\documentclass[runningheads]{llncs}
\usepackage[T1]{fontenc}
\usepackage{graphicx}
\usepackage{latexsym}
\usepackage{microtype}
\usepackage{amsfonts}
\usepackage{graphicx}
\usepackage{amsmath}
\usepackage{booktabs}
\usepackage[misc]{ifsym}
\usepackage[colorlinks,linkcolor=blue]{hyperref}

\begin{document}
\title{Document-Level Relation Extraction with Relation Correlation Enhancement}
\titlerunning{LACE for Document-Level Relation Extraction}
\author{Yusheng Huang, Zhouhan Lin\thanks{Zhouhan Lin is the corresponding author.}}
\authorrunning{Y. Huang, and Z. Lin}
%
\institute{Shanghai Jiao Tong University, Shanghai, China \\
\email{huangyusheng@sjtu.edu.cn, lin.zhouhan@gmail.com}}
\maketitle              
\begin{abstract}
Document-level relation extraction (DocRE) is a task that focuses on identifying relations between entities within a document. However, existing DocRE models often overlook the correlation between relations and lack a quantitative analysis of relation correlations. To address this limitation and effectively capture relation correlations in DocRE, we propose a relation graph method, which aims to explicitly exploit the interdependency among relations. Firstly, we construct a relation graph that models relation correlations using statistical co-occurrence information derived from prior relation knowledge. Secondly, we employ a re-weighting scheme to create an effective relation correlation matrix to guide the propagation of relation information. Furthermore, we leverage graph attention networks to aggregate relation embeddings. Importantly, our method can be seamlessly integrated as a plug-and-play module into existing models. Experimental results demonstrate that our approach can enhance the performance of multi-relation extraction, highlighting the effectiveness of considering relation correlations in DocRE. \footnote{Codes are available at \href{https://github.com/LUMIA-Group/LACE}{https://github.com/LUMIA-Group/LACE}}

\end{abstract}

\section{Introduction}
Relation extraction (RE) plays a vital role in information extraction by identifying semantic relations between target entities in a given text. Previous research has primarily focused on sentence-level relation extraction, aiming to predict relations within a single sentence \cite{DBLP:conf/aaai/FengHZYZ18}. However, in real-world scenarios, valuable relational facts are often expressed through multiple mentions scattered across sentences, such as in Wikipedia articles \cite{DBLP:journals/tacl/PengPQTY17}. Consequently, the extraction of relations from multiple sentences, known as document-level relation extraction, has attracted significant research attention in recent years.

\begin{figure}[t]
\centering
\includegraphics[width=1.00\textwidth]{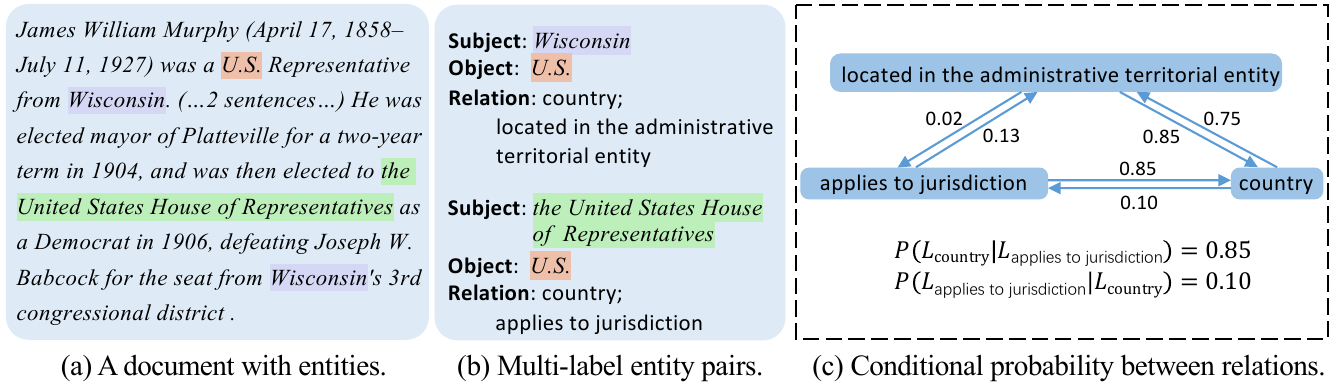}
\caption{Examples of relation correlation for multi-relation extraction. (a) presents a document containing multiple entities. (b) illustrates the multi-relation entity pairs. For instance, the subject entity \emph{Wisconsin} and the object entity \emph{U.S.} express the \emph{country} and \emph{located in the administrative territorial entity} relations. (c) demonstrates the conditional probabilities between three relations, which are derived from the DocRED dataset.}
\label{graph1_example}
\end{figure}

Compared to sentence-level RE, document-level RE presents unique challenges in designing model structures. In sentence-level RE, a single relation type is associated with each entity pair, as observed in SemEval 2010 Task 8 \cite{DBLP:conf/naacl/HendrickxKKNSPP09} and TACRED \cite{DBLP:conf/emnlp/ZhangZCAM17}. However, in document-level RE, an entity pair can be associated with multiple relations, making it more challenging than sentence-level RE. Figure \ref{graph1_example}(b) illustrates multi-relation examples extracted from the DocRED dataset \cite{DBLP:conf/acl/YaoYLHLLLHZS19}, where each entity pair is associated with two distinct relations. Moreover, in document-level RE, the number of relation types to be classified can be large (e.g., 97 in the DocRED dataset), further increasing the difficulty of extracting multiple relations.

To address this challenge, previous studies have commonly approached it as a multi-label classification problem, where each relation is treated as a label. Binary cross-entropy loss is typically employed to handle this multi-label scenario \cite{DBLP:conf/acl/NanGSL20,DBLP:conf/emnlp/ZengXCL20}. During inference, a global threshold is applied to determine the relations. More recently, \cite{DBLP:conf/acl/LiXLFRJ21} utilize the asymmetric loss (ASL) \cite{DBLP:journals/corr/abs-2009-14119} to mitigate the imbalance between positive and negative classes. Additionally, \cite{DBLP:conf/ijcai/ZhangCXDTCHSC21} propose to employ a balanced softmax method to mitigate the imbalanced relation distribution, where many entity pairs have no relation. \cite{DBLP:conf/aaai/Zhou0M021} introduce the adaptive thresholding technique, which replaces the global threshold with a learnable threshold class. However, previous studies have rarely quantitatively analyzed the co-occurred relations and have not explicitly utilized this feature.

According to the statistics in DocRED dataset, we find that relations co-occur with priors. As illustrated in Figure \ref{graph1_example}(c), for entity pairs with multiple relations, the conditional probability of relation \emph{country} appears given that relation \emph{applies to jurisdiction} appears is $0.85$, while the conditional probability of relation \emph{applies to jurisdiction} appears given that relation \emph{country} appears is $0.10$. Besides, with great chance, relation \emph{country} and relation \emph{located in the administrative territorial entity} appear together. Considering the relations exhibit combinatorial characteristics, it is desirable to employ the relation correlations to ameliorate the model structure and boost the multi-relation extraction. 

In this paper, we aim to tackle the challenge of multi-relation extraction in document-level RE by leveraging the correlation characteristics among relations. Specifically, we propose a relation graph method that leverages the prior knowledge of interdependency between relations to effectively guide the extraction of multiple relations. To model the relation correlations, we estimate it by calculating the frequency of relation co-occurrences in the training set \cite{DBLP:conf/aaai/WangHLLZMW20,DBLP:journals/fss/CheCM22}. To avoid overfitting, we filter out noisy edges below a certain threshold and create a conditional probability matrix by dividing each co-occurrence element by the occurrence numbers of each relation. This matrix is then binarized to enhance the model's generalization capability, and the relation graph is constructed as a binary directed graph. Additionally, we employ a re-weighting scheme to construct an effective relation correlation matrix, which guides the propagation of relation information \cite{DBLP:conf/cvpr/ChenWWG19}. We employ Graph Attention Networks (GAT) \cite{DBLP:conf/iclr/VelickovicCCRLB18} with the multi-head graph attention to aggregate relation embeddings. Based on the adaptive thresholding technique \cite{DBLP:conf/aaai/Zhou0M021}, the loss function in our method is also amended by emphasizing the multi-relation logits. Our method is easy for adoption as it could work as a plug-in for existing models. We conduct extensive experiments on the widely-used document-level RE dataset DocRED, which contains around $7\%$ multi-relation entity pairs. Experimental results demonstrate the effectiveness of our method, achieving superior performance compared to baseline models. In summary, our contributions are as follows:
\begin{itemize}
    \item We conduct comprehensive quantitative studies on relation correlations in document-level RE,  providing insights for addressing the challenge of multi-relation extraction.
    \item We propose a relation graph method that explicitly leverages relation correlations, offering a plug-in solution for other effective models.
    \item We evaluate our method on a large-scale DocRE dataset, demonstrating its superior performance compared to baselines.
\end{itemize}

\section{Related Work}\label{relatedwork}
Relation extraction, a crucial task in natural language processing, aims to predict the relations between two entities. It has widespread applications, including dialogue generation \cite{DBLP:conf/acl/HeBEL17} and question answering \cite{DBLP:conf/naacl/HixonCH15}. Previous researches largely focus on sentence-level RE, where two entities are within a sentence. Many models have been proposed to tackle the sentence-level RE task, encompassing various blocks such as CNN \cite{DBLP:conf/coling/ZengLLZZ14,DBLP:conf/acl/SantosXZ15}, LSTM \cite{DBLP:conf/acl/ZhouSTQLHX16,DBLP:conf/acl/CaiZW16}, attention mechanism \cite{DBLP:conf/acl/WangCML16,DBLP:conf/naacl/YeL19}, GNN \cite{DBLP:conf/acl/GuoZL19,DBLP:conf/acl/ZhuLLFCS19}, and transformer \cite{DBLP:conf/aaai/XiaoTF0Z20,DBLP:conf/acl-clinicalnlp/ChenLDL20}. 

Recent researches work on document-level relation extraction since many real-world relations can only be extracted from multiple sentences \cite{DBLP:conf/acl/YaoYLHLLLHZS19}. From the perspective of techniques, various related approaches could be divided into the graph-based category and the transformer-based category. For the graph-based models that are advantageous to relational reasoning, \cite{DBLP:conf/acl/NanGSL20} propose LSR that empowers the relational reasoning across multiple sentences through automatically inducing the latent document-level graph. \cite{DBLP:conf/emnlp/ZengXCL20} propose GAIN with two constructed graphs that captures complex interaction among mentions and entities. \cite{DBLP:conf/coling/ZhouXYLLJ20} propose GCGCN to model the complicated semantic interactions among multiple entities. \cite{DBLP:conf/coling/LiYSXXZ20} propose to characterize the complex interaction between multiple sentences and the possible relation instances via GEDA networks. \cite{DBLP:conf/coling/ZhangYSLTWG20} introduce DHG for document-level RE to promote the multi-hop reasoning. For the transformer-based models that are capable of implicitly model long-distance dependencies, \cite{DBLP:journals/corr/abs-1909-11898} discover that using pre-trained language models can improve the performance of this task. \cite{DBLP:conf/pakdd/TangC0CFWY20} propose HIN to make full use of the abundant information from entity level, sentence level and document level. \cite{DBLP:conf/emnlp/YeLDLLSL20} present CorefBERT to capture the coreferential relations in context. Recent works normally directly leverage the pre-trained language models such as BERT \cite{DBLP:conf/naacl/DevlinCLT19} or RoBERTa \cite{DBLP:journals/corr/abs-1907-11692} as word embeddings. 

\section{Methodology}
In this section, we provide a detailed explanation of our LAbel Correlation Enhanced (LACE) method, for document-level relation extraction. We begin by formulating the task in $\S\ref{m1}$ and then introduce the overall architecture in $\S\ref{m2}$. In $\S\ref{m3}$, we discuss the encoder module for obtaining the feature vectors of entity pairs. The relation correlation module, outlined in $\S\ref{m4}$, is designed to capture relation correlations. Finally, we present the classification module with multi-relation adaptive thresholding loss for model optimization in $\S\ref{m5}$.

\subsection{Task Formulation}\label{m1}
Given an input document that consists of $N$ entities $\mathcal{E}=\left\{e_{i}\right\}_{i=1}^{N}$, this task aims to identity a subset of relations from $\mathcal{R} \cup \{\mathrm{NA}\}$ for each entity pair $\left(e_{s}, e_{o}\right)$, where $s, o = 1, ..., N; s\neq o$. The first entity $e_s$ is identified as the \emph{subject} entity and the second entity $e_o$ is identified as the \emph{object} entity. $\mathcal{R}$ is a pre-defined relation type set, and $\mathrm{NA}$ denotes no relation expressed for the entity pair. Specifically, an entity $e_i$ can contain multiple mentions with different surface names $e_i^k, k = 1, ..., m$. During testing, the trained model is supposed to predict labels of all the entity pairs $\left(e_{s}, e_{o}\right)_{s, o = 1, ..., N; s\neq o}$ within documents.

\subsection{Overall Architecture}\label{m2}
\begin{figure*}[t]
\centering
\includegraphics[width=0.83\textwidth]{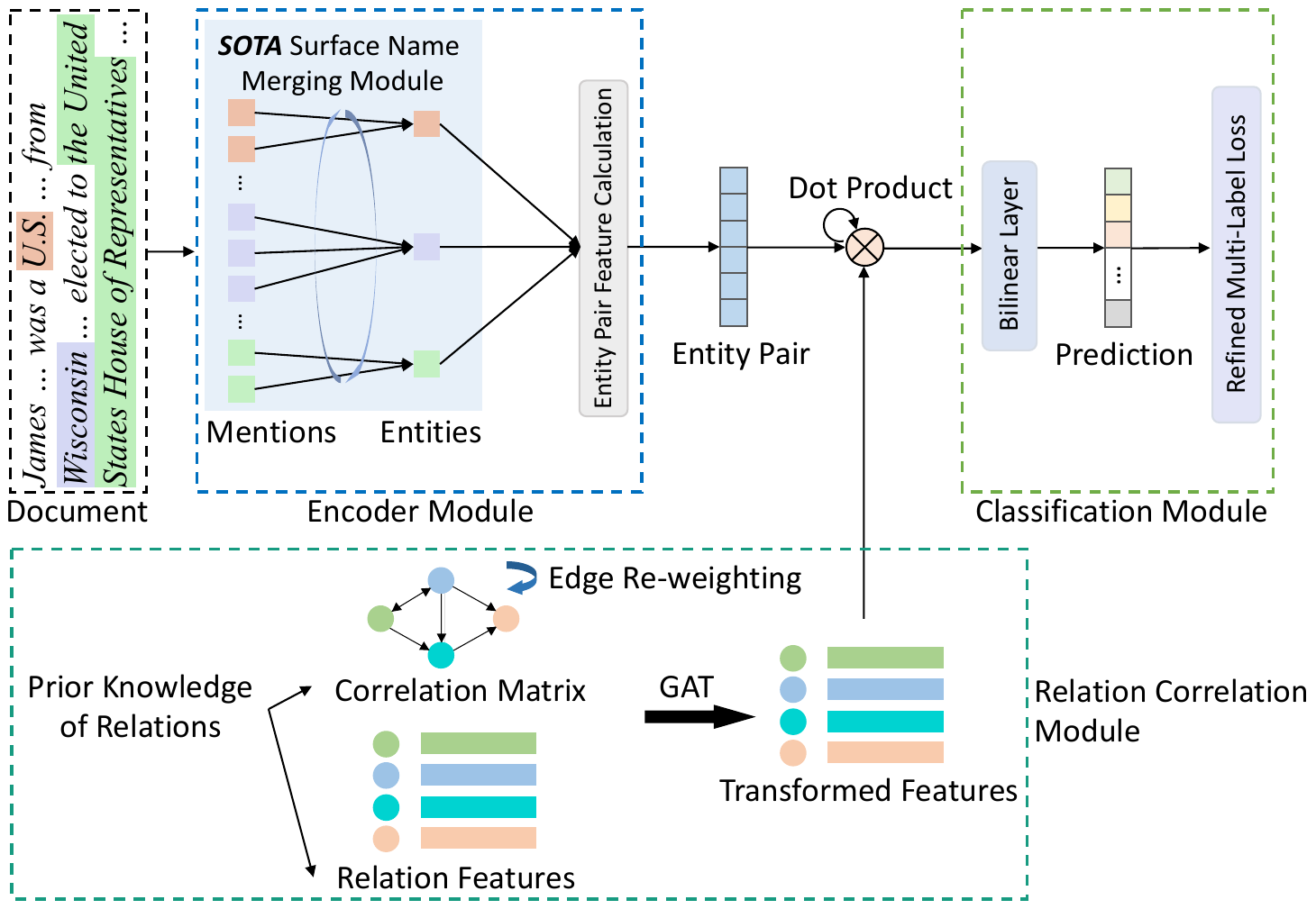}
\caption{Architecture of our LACE method, consisting of three modules: the encoder module, relation correlation module, and the classification module. }
\label{framework}
\end{figure*}

As illustrated in Figure \ref{framework}, the overall architecture consists of three modules. The encoder module first yields the contextual embeddings of all the entity mentions, and then each entity embedding is obtained by integrating information from the corresponding entity mentions, i.e. surface name merging. Afterward, entity pair features are calculated to enhance the entity pair embedding. The relation correlation module generates relation feature vectors. The correlation matrix is built in a data-driven manner, which is based on the statistics of the provided training set. We employ the edge re-weighting scheme to create a weighted adjacency matrix, which is beneficial for deploying graph neural networks. GAT is applied to the correlation matrix and relation features to generate more informative relation feature vectors. In the classification module, a bi-linear layer is utilized for prediction. Besides, based on the adaptive thresholding technique, we resort to a refined loss function for better multi-label classification. 

\subsection{Encoder Module}\label{m3}
For a document consisting of $l$ words $\left[x_{t}\right]_{t=1}^{l}$ and $N$ entities (each entity containing several mentions), we obtain the word embedding $x_{i}^{w}$ and entity type embedding $x_{i}^{t}$. Then we concatenate them and feed them to BiLSTM layers to generate the contextualized input representations $\boldsymbol{h}_{i}$:
\begin{equation}
   \boldsymbol{h}_{i} = \operatorname{BiLSTM}([x_{i}^{w}; x_{i}^{t}]).
\end{equation}
Mention representations $m_i$ are obtained by conducting a max-pooling operation on the words, and entity representations $\boldsymbol{E}_{i}$ are generated by the log-sum-exp pooling over all the entity mention representations $m_i$: 
\begin{equation}
\boldsymbol{E}_{i}=\log \sum_{i=1}^{j} \exp \left(m_i\right),
\end{equation}
where $j$ is the number of entity mentions.

In this way, we can generate the embeddings of the head entity and tail entity, denoted as $\boldsymbol{E}_{s}\in \mathbb{R}^{d_B}$ and $\boldsymbol{E}_{o}\in \mathbb{R}^{d_B}$, respectively. The entity pair features are obtained by concatenating these embeddings.

Many studies fall in this module, which focus on generating more contextual representations by leveraging Transformer \cite{DBLP:conf/aaai/Zhou0M021} or document graphs  \cite{DBLP:journals/corr/abs-2009-14119}. 
These studies can be seamlessly integrated into our LACE method, or conversely, LACE can be incorporated as a plug-in for other models.

\subsection{Relation Correlation Module}\label{m4}
We model the relation correlation interdependency in the form of conditional probability, i.e., $P\left(L_{b} \mid L_{a}\right)$ means the probability of occurrence of relation type $L_b$ when relation type $L_a$ appears. Considering that for conditional probabilities $P\left(L_{b} \mid L_{a}\right) \neq P\left(L_{a} \mid L_{b}\right)$, we construct a relation-related directed graph based on the relation prior knowledge of the training set for modeling, which means that the adjacency matrix is asymmetric. 

\subsubsection{Correlation Matrix Construction}
To construct the correlation matrix, we first count the co-occurrence of relations in the training set and obtain the co-occurrence matrix $\boldsymbol{C}^{r\times r}$, where $r$ is the number of pre-defined relation types. To obtain the conditional probabilities between relations, each element in the co-occurrence matrix $\boldsymbol{C}^{r\times r}$ is divided by the total number of relation co-occurrences, i.e., 
\begin{equation}
P_{ij}=C_{ij} / \sum_{j} C_{ij},
\end{equation}
where $P_{ij}=P\left(L_{j} \mid L_{i}\right)$ denotes the probability of relation type $L_j$ when relation type $L_i$ appears. 

However, the above method for correlation matrix construction may suffer two drawbacks. Firstly, some relations rarely appear together with others. This will lead to a large probability value for the co-occurred relation, which is unreasonable. Secondly, there may be a deviation between the statistics of the training dataset and the statistics of the test dataset. Using the exact numbers tend to overfit the training dataset, which might hurt the generalization capacity. Therefore, to alleviate these issues, we set a threshold $\tau$ to filter these rare co-occurred relations. Then we binarize the conditional probability matrix $\boldsymbol{P}$ by
\begin{equation}\label{eq7}
{B}_{i j}= \begin{cases}0, & \text { if } {P}_{i j}<\delta \\ 1, & \text { if } {P}_{i j} \geq \delta\end{cases},
\end{equation}
where $\boldsymbol{B}$ is the binarized correlation matrix. $\delta$ is the conditional probability threshold. Besides, We add the self-loop by setting $B_{ii}=1, i = 1, ..., r$. 
\subsubsection{Edge Re-weighting Scheme}
One concern for utilizing the binary correlation matrix $\boldsymbol{B}$ for graph neural networks is the over-smoothing issue \cite{DBLP:journals/aiopen/ZhouCHZYLWLS20} that the node attribute vectors tend to converge to similar values. There is no natural weight difference between the relation features and its neighbor nodes'. To mitigate this issue, we employ the following re-weighting scheme, 
\begin{equation}\label{eq8}
R_{ij}=\left\{\begin{array}{ll}
p / \sum_{j=1 \atop i \neq j}^{r} B_{ij}, & \text { if } i \neq j \\
1-p, & \text { if } i=j
\end{array},\right.
\end{equation}
where $\boldsymbol{R}$ is the re-weighted relation correlation matrix and $p$ is a hyper-parameter. In this way, the fixed weights for the relation feature and its neighbors will be applied during training, which is beneficial for alleviating this issue. 

Relation features are the embedding vectors obtained in the same way as word embeddings. We then exploit GAT networks with a K-head attention mechanism to aggregate relation features for two reasons. First, GAT is suitable for directed graphs. Second, GAT maintains a stronger representation ability since the weights of each node can be different. The transformed features by GAT are denoted as $\boldsymbol{R} \in \mathbb{R}^{r\times d_B}$. 

\subsection{Classification Module}\label{m5}
Given feature vectors $\boldsymbol{E}_s, \boldsymbol{E}_o \in \mathbb{R}^{d_B}$ of the entity pair $(e_s, e_o)$ and the transformed relation features $\boldsymbol{R} \in \mathbb{R}^{r\times d_B}$, we map them to hidden representations $\boldsymbol{I_s}, \boldsymbol{I_o}\in \mathbb{R}^r$ followed by the layer normalization operation, 
\begin{equation}
\boldsymbol{I_s}=\text { LayerNorm }\left( \boldsymbol{R} \cdot \boldsymbol{E}_s \right),
\end{equation}
\begin{equation}
\boldsymbol{I_o}=\text { LayerNorm }\left( \boldsymbol{R} \cdot \boldsymbol{E}_o \right).
\end{equation}
Then, we obtain the prediction probability of the relation $r^{\prime}$ via a bilinear layer, 
\begin{equation}\label{equation1}
\mathrm{P}_{r^{\prime}}=\sigma\left((\boldsymbol{E}_s\oplus \boldsymbol{I_s})^\top W_{r} (\boldsymbol{E}_o\oplus \boldsymbol{I_o}) +b_{r}\right),
\end{equation}
where $\sigma$ is the sigmoid activation function. $W_{r} \in \mathbb{R}^{(d_B+r) \times (d_B+r)}$, $b_{r} \in \mathbb{R}$ are model parameters, and $\oplus$ denotes the concatenation operation. 

Previous study \cite{DBLP:conf/aaai/Zhou0M021} has shown the effectiveness of the Adaptive Thresholding loss (AT loss), where a threshold class is set such that logits of the positive classes are greater than the threshold class while the logits of the negative classes are less than the threshold class. However, their designed loss function does not quite match the multi-label problem, since they implicitly use the softmax function in the calculation of the positive-class loss function.
Therefore, during each loss calculation, the AT loss is unable to extract multiple relations. The superposition of multiple calculations would result in a significant increase in time overhead. To mitigate this issue, we propose a novel loss function called Multi-relation Adaptive Thresholding loss (MAT loss), which is defined as follows, 
\begin{equation}
\mathcal{L}_{+}= - \log ( 1- \mathrm{P}({\tt TH}) ) - \sum_{r^{\prime}\in L_p}( y^{r^{\prime}} \log \mathrm{P}(r^{\prime}) + \left(1-y^{r^{\prime}}\right) \log ( 1- \mathrm{P}(r^{\prime}) )),
\end{equation}
\begin{equation}
\mathcal{L}_{-}=-\log \left(\frac{\exp \left(\operatorname{logit}_{{\tt TH}}\right)}{\sum_{r^{\prime} \in L_o \cup\{{\tt TH}\}} \exp \left(\operatorname{logit}_{r^{\prime}}\right)}\right),
\end{equation}
where the threshold class ${\tt TH}$ is the ${\mathrm{NA}}$ class. $L_p$ and $L_o$ denote the relations the exist and do not exist between the entity pair, respectively. ${\rm logit}$ means the number without $\sigma$ in Equation \ref{equation1}. The final loss function is $\mathcal{L} = \alpha\mathcal{L}_{+} + (1-\alpha)\mathcal{L}_{-}$, where $\alpha$ is a hyper-parameter. In this way, our MAT loss enables the extraction of multiple relations.

During inference, we assign labels to entity pairs whose prediction probabilities meet the following criteria, 
\begin{equation} \label{eq15}
\mathrm{P}\left(r^{\prime} \mid e_{s}, e_{o}\right) \ge (1+ \theta) \ \mathrm{P}({\tt TH}),
\end{equation}
where $\theta$ is a hyper-parameter that maximizes evaluation metrics. 

\section{Experiments}
\subsection{Dataset}
We evaluate our proposed approach on a large-scale human-annotated dataset for document-level relation extraction DocRED \cite{DBLP:conf/acl/YaoYLHLLLHZS19}, which is constructed from Wikipedia articles. DocRED is larger than other existing counterpart datasets in aspects of the number of documents, relation types, and relation facts. Specifically, DocRED contains 3053 documents for the training set, 1000 documents for the development set, and 1000 documents for the test set, with 96 relation types and 56354 relational facts. For entity pairs with relations, around $7\%$ of them express more than one relation type, and an entity pair can express up to $4$ relations. 
\footnote{We conduct no experiments on the CDR \cite{DBLP:journals/biodb/LiSJSWLDMWL16} and GDA \cite{DBLP:conf/recomb/WuLLTL19} datasets in the biomedical domain, because they do not suffer the multi-relation issue. Therefore, they do not match our scenario.}

\subsection{Implementation Details}
We employ GloVe \cite{DBLP:conf/emnlp/PenningtonSM14} and BERT-based-cased \cite{DBLP:conf/naacl/DevlinCLT19} word embeddings in the encoder module, respectively. When employing GloVe word embeddings, we use Adam optimizer with learning rate being $e^{-3}$. When employing BERT-based-cased, we use AdamW with a linear warmup for the first 6\% steps. The learning rate for BERT parameters is $5e^{-5}$ and $e^{-4}$ for other layers. In the relation correlation module, we set the threshold $\tau$ to be 10 for filtering noisy co-occurred relations, and $\delta$ is set to be 0.05 in Equation \ref{eq7}. We set $p$ to be 0.3 in Equation \ref{eq8} and $\theta$ to be 0.85 in Equation \ref{eq15}. We employ 2-layer GAT networks with $k=2$ attention heads computing 500 hidden features per head. We utilize the exponential linear unit (ELU) as the activation function between GAT layers. $\alpha$ in the classification module is 0.4. All hyper-parameters are tuned on the development set. 

\subsection{Baseline Systems}
We compare our approach with the following models, including three categories. 

\paragraph{GloVe-based Models.}
These models report results using GloVe word embeddings and utilize various neural network architectures including CNN, BiLSTM, and Context-Aware \cite{DBLP:conf/acl/YaoYLHLLLHZS19}, to encode the entire document, and then obtain the embeddings of entity pairs for relation classification. The recent mention-based-reasoning model MRN \cite{DBLP:journals/corr/abs-2009-14119} also present the results with GloVe word embedding.

\paragraph{Transformer-based Models.} These models directly exploit the pre-trained language model BERT for document encoding without document graph construction, including HIN-BERT \cite{DBLP:conf/pakdd/TangC0CFWY20}, CorefBERT \cite{DBLP:conf/emnlp/YeLDLLSL20}, and ATLOP-BERT \cite{DBLP:conf/aaai/Zhou0M021}. We mainly compare our method LACE with ATLOP model that aims to mitigate the multi-relation problem.
\paragraph{Graph-based Models.} Homogeneous or heterogeneous graphs are constructed based on the document features for reasoning. Then, various graph-based models are leveraged to perform inference on entity pairs, including BiLSTM-AGGCN \cite{DBLP:conf/acl/GuoZL19}, LSR-BERT \cite{DBLP:conf/acl/NanGSL20}, GAIN-BERT \cite{DBLP:conf/emnlp/ZengXCL20}. The MRN-BERT \cite{DBLP:journals/corr/abs-2009-14119} aims to capture the local and global interactions via multi-hop mention-level reasoning.

When compared to GloVe-based models and graph-based models, we integrate the MRL layer from MRN into the encoder module. When compared to Transformer-based models, we incorporate the localized context pooling technique from ATLOP into the encoder module.

\begin{table}
\caption{Results on the development set and test set of DocRED. }
\centering
\begin{tabular}{lcccc}
\toprule
\textbf{Model}  & \multicolumn{2}{c}{\textbf{Dev}} &   \multicolumn{2}{c}{\textbf{Test}} \\
    & Ign $F_1$ & $F_1$ & Ign $F_1$ & $F_1$ \\
\midrule
\emph{With GloVe} \\
CNN \cite{DBLP:conf/acl/YaoYLHLLLHZS19} & 41.58 & 43.45 & 40.33 & 42.26 \\
BiLSTM \cite{DBLP:conf/acl/YaoYLHLLLHZS19} & 48.87 & 50.94 & 48.78 & 51.06 \\
Context-Aware \cite{DBLP:conf/acl/YaoYLHLLLHZS19} & 48.94 & 51.09 & 48.40 & 50.70 \\
MRN \cite{DBLP:conf/acl/LiXLFRJ21} & 56.62 & 58.59 & 56.19 & 58.46 \\
\midrule
\textbf{LACE} & \textbf{57.01} & \textbf{58.92} & \textbf{56.61} & \textbf{58.64} \\
\midrule
\midrule
\emph{With BERT+Transformer} \\ 
HIN-BERT \cite{DBLP:conf/pakdd/TangC0CFWY20} & 54.29  & 56.31 &  53.70 &  55.60  \\
CorefBERT \cite{DBLP:conf/emnlp/YeLDLLSL20} & 55.32  & 57.51 & 54.54  & 56.96 \\
ATLOP-BERT \cite{DBLP:conf/aaai/Zhou0M021} & 59.22  & 61.09 & 59.31 &  61.30 \\
\midrule
\textbf{LACE-BERT} & \textbf{59.58} & \textbf{61.43} & \textbf{59.40} & \textbf{61.50} \\
\midrule
\emph{With BERT+Graph} \\
BiLSTM-AGGCN \cite{DBLP:conf/acl/GuoZL19}  &  46.29 &  52.47 & 48.89 &  51.45 \\
LSR-BERT \cite{DBLP:conf/acl/NanGSL20} & 52.43 &  59.00 & 56.97 &  59.05 \\
GAIN-BERT \cite{DBLP:conf/emnlp/ZengXCL20} &  59.14 &  61.22 & 59.00 &  61.24 \\
MRN-BERT \cite{DBLP:conf/acl/LiXLFRJ21} &  59.74 &  61.61 & 59.52 &  61.74 \\
\midrule
\textbf{LACE-MRL-BERT} & \textbf{59.98} & \textbf{61.75} & \textbf{59.85} & \textbf{61.90} \\
\bottomrule
\end{tabular}
\label{result}
\end{table}

\subsection{Quantitative Results}
Table \ref{result} shows the experimental results on the DocRED dataset. 
Following previous studies \cite{DBLP:conf/acl/YaoYLHLLLHZS19,DBLP:conf/aaai/Zhou0M021}, we adopt the Ign F$_1$ and F$_1$ as the evaluation metrics, where Ign F1 is calculated by excluding the shared relation facts between the training set and development/test set. 

For the GloVe-based models, our method LACE achieves 56.61\% Ign F$_1$ and 58.64\% F$_1$-score on the test set, outperforming all other methods. For the transformer-based models using BERT, our method LACE-BERT achieves $61.50\%$ F1-score on the test set, which outperforms the ATLOP-BERT model. These experimental results also show that the pre-trained language model can cooperate well with the LACE method. For the graph-based models, we chieve 61.90\% F$_1$-score on the test set. 
The result demonstrates that capturing the mention-level contextual information is helpful and our proposed method could work well with the mention-based reasoning method.
Overall, results demonstrate the effectiveness of leveraging the relation information. 

\subsection{Analysis of Relation Correlation Module}
We investigate the effect of key components in the relation correlation module.
\paragraph{Matrix Construction Threshold.} As shown in Table \ref{threshold}, we analyze the effect of probability filtering threshold $\delta$ in Equation \ref{eq7}. We obtain the highest F$_1$ score when $\delta$ equals 0.05 for all experiments. Besides, results indicates that $\delta=0.03$ will lead to more performance degradation compared with $\delta=0.07$. We believe that this is due to the smaller threshold value resulting in more noise edges.

\begin{table}
\centering
\caption{F1-score on the development set when tuning the probability filtering threshold $\delta$.}
\begin{tabular}{lccc}
\toprule
\textbf{Model}  & 3\% & 5\% & 7\% \\
\midrule
LACE    &  58.74 & \textbf{58.92} & 58.82 \\
LACE-BERT   & 61.38  & \textbf{61.43} &  61.40 \\
LACE-MRL-BERT    & 61.67  & \textbf{61.75} & 61.71 \\
\bottomrule
\end{tabular}
\label{threshold}
\end{table}

\begin{table}
\caption{F1-score on the development set with different GAT layers. $L$ denotes layer.}
\centering
\begin{tabular}{lccc}
\toprule
\textbf{Model}  & 1-$L$ & 2-$L$ & 3-$L$ \\
\midrule
LACE    &  58.80 & \textbf{58.92} & 58.64 \\
LACE-BERT   & 61.40  & \textbf{61.43} &  61.23 \\
LACE-MRL-BERT    & 61.69  & \textbf{61.75} & 61.63 \\
\bottomrule
\end{tabular}
\label{GAT}
\end{table}

\paragraph{GAT layer.}
We report the results of different GAT layers with two heads in Table \ref{GAT}. 
Results demonstrates that 1-layer and 2-layer GAT networks achieves relatively similar results, while 3-layer GAT networks leads to greater performance degradation. 
The probable reason for the performance degradation might be the over-smoothing issue, that is, the node feature vectors are inclined to converge to comparable values.

\begin{table}
\caption{F1-score for multi-relation extraction on the development set. \emph{Rel} denotes relations.}
\centering
\begin{tabular}{lccc}
\toprule
\textbf{Model}  & 2-\emph{Rel} & 3-\emph{Rel} & Overall \\
\midrule
ATLOP-BERT    & 40.13  & 29.59 &  39.62   \\
LACE-BERT       & \textbf{42.03}  & \textbf{32.58} & \textbf{41.55}  \\
\bottomrule
\end{tabular}
\label{multilabelperformance}
\end{table}

\subsection{Performance on Multi-Label Extraction}
In order to evaluate the performance of multi-label extraction, we re-implement ATLOP-BERT model and report the experimental results of multi-relation extraction as shown in Table \ref{multilabelperformance}. As seen, our approach LACE-BERT gains 1.9\% and 2.99\% F1-score improvement on the 2-relation and 3-relation extraction, respectively, which demonstrates the effectiveness of leveraging the relation correlations. Overall, our approach achieves 1.93\% F1-score improvements on multi-label extraction compared with ATLOP-BERT. 

\begin{table}
\caption{F1-score on the development set for ablation study. RCM denotes the relation correlation module.}
\centering
\begin{tabular}{lccc}
\toprule
\textbf{Model}  & 2-\emph{Relation} & 3-\emph{Relation} & Overall \\
\midrule
LACE-BERT        & \textbf{42.03}  & \textbf{32.58} & \textbf{41.55} \\
\midrule
- RCM & 40.74 & 30.62 & 40.25 \\
- $\mathcal{L}_{MAT}$ & 41.43 & 31.78 & 40.94 \\
\bottomrule
\end{tabular}
\label{ablation}
\end{table}

\subsection{Ablation Study}
We conduct ablation studies to verify the necessity of two critical modules in LACE-BERT for multi-relation extraction as depicted in Table \ref{ablation}. Results show that two modules contribute to the final improvements. Firstly, removing the relation correlation module causes more performance degradation, and we thus believe that leveraging the prior knowledge of relation interdependency is helpful for the multi-relation extraction. Secondly, we replace our multi-relation adaptive thresholding loss $\mathcal{L}_{MAT}$ with the adaptive thresholding loss \cite{DBLP:conf/aaai/Zhou0M021} for comparison. We believe that the reason for the improvement is that the MAT loss enlarges the margin values between all the positive classes and the threshold class.

\begin{figure*}[t]
\centering
\includegraphics[width=1.00\textwidth]{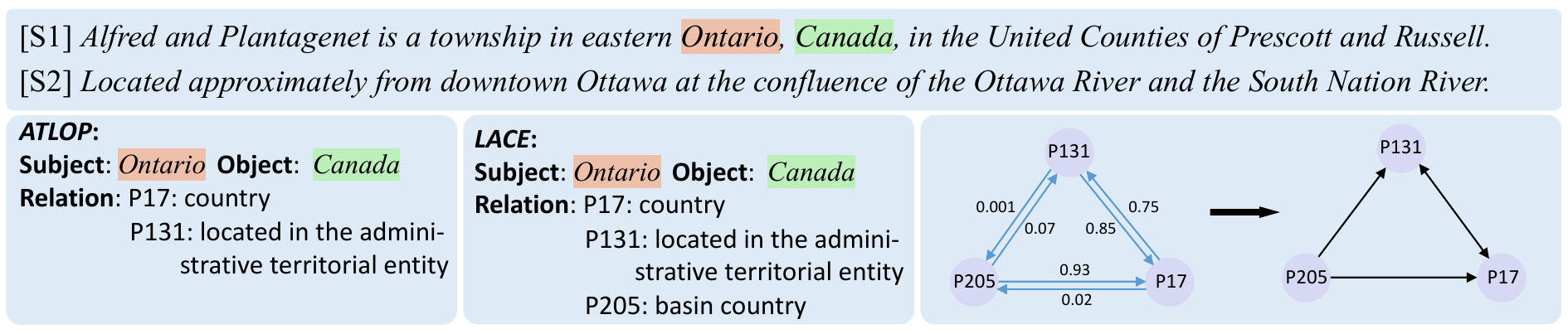}
\caption{Case study of a triple-relation entity pair from the development set of DocRED. We visualize the conditional probabilities among these relations and exhibit the constructed directed sub-graph.}
\label{case}
\end{figure*}

\subsection{Case Study}
Figure \ref{case} shows a case study of our proposed approach LACE-BERT, in comparison with ATLOP-BERT baseline. We can observe that ATLOP-BERT can only identify the \emph{P17} and \emph{P131} relations for the entity pair \emph{(Ontario, Canada)}, where the two relations frequently appear together. However, ATLOP-BERT fails to identity the \emph{P205} relation, while LACE-BERT deduces this relation. By introducing the label correlation matrix, this relation \emph{P205} establishes connections with other relations with high conditional probabilities, which is advantageous for multi-relation extraction. 

\section{Conclusion}
In this work, we propose our method LACE for document-level relation extraction. LACE includes a relation graph construction approach which explicitly leverages the statistical co-occurrence information of relations. Our method effectively captures the interdependency among relations, resulting in improved performance on multi-relation extraction. Experimental results demonstrate the superior performance of our proposed approach on a large-scale document-level relation extraction dataset.

\subsubsection{Acknowledgements}
The  authors  would  like  to  thank  the  support  from the National Natural Science Foundation of China (NSFC) grant (No. 62106143), and Shanghai Pujiang Program (No. 21PJ1405700).

%
%
%
\bibliographystyle{splncs04}
\bibliography{ref}

\end{document}